\documentstyle[11pt]{article}

    \advance\textheight by \topskip

\begin{document}

\hfill UM-TH-97-03

\hfill NIKHEF-97-012

\begin{center}
{\huge \bf The 4-loop quark mass anomalous dimension
and the invariant quark mass.} \\[8mm]
J.A.M. Vermaseren$^a$,  S.A. Larin$^{b}$, T. van Ritbergen$^c$ \\ [3mm]
\begin{itemize}
\item[$^a$]
 NIKHEF, P.O. Box 41882, \\ 1009 DB, Amsterdam, The Netherlands \\
\item[$^b$]
 Theory Division, CERN, CH -1211, Geneva 23, Switzerland\\
 and INR of the Russian Academy of Sciences,   \\
 60th October Anniversary Prospect 7a,
 Moscow 117312, Russia
\item[$^c$]
 Randall Laboratory of Physics, University of Michigan,\\
 Ann Arbor, MI 48109, USA
\end{itemize}
\end{center}

\begin{abstract}
We present the analytical calculation of the four-loop
quark mass anomalous dimension in
Quantum Chromodynamics within the minimal subtraction scheme.
On the basis of this result we find that 
the so-called invariant quark mass is a very good reference mass 
for the accurate evolution
of the running $\overline{\rm MS}$ quark mass in phenomenological
applications. We also obtain for the first time
a complete 4-th order perturbative QCD expression for a physical quantity,
the total Higgs boson decay rate into hadrons, and analyze the infrared
fixed point for this case.
\vspace{3cm} \\
PACS numbers: 13.85.Hd, 12.38.-t, 12.38.Bx
\end{abstract}
\newpage


The renormalization group equations play an important role in 
our understanding of Quantum Chromodynamics and the strong
interactions.
The beta function and the quarks mass anomalous dimension are among 
the most prominent objects that appear in these renormalization group equations.

The quark mass anomalous dimension in QCD was calculated 
at the 3-loop level in \cite{loop3}.
In a previous paper \cite{beta4loop} we calculated analytically
the 4-loop QCD $\beta$-function.
Here we complete this 4-loop project by obtaining the quark mass
anomalous dimension in the 4-loop order.

We will apply the obtained anomalous dimension to show the relevance 
of the invariant quark mass for an accurate evolution of the running
quark masses in phenomenological applications. 
We also apply the 4-loop result to analyze a possible QCD infrared
fixed point in the case of Higgs boson decay into hadrons.

Throughout the calculations in this article we use the technique of 
Dimensional Regularization
\cite{hv} for the regularization of ultraviolet divergences
and the Minimal Subtraction (MS) scheme \cite{h}
or its standard modification, the $\overline{\rm MS}$ scheme
\cite{bbdm}, for renormalization.
The dimension of space-time is defined as $D=4-2 \varepsilon$, where
$\varepsilon$ is the regularzation parameter fixing the deviation of the
space-time dimension from its physical value 4.

Let us write the full Lagrangian of Quantum Chromodynamics
with massive quarks in the covariant gauge
\begin{eqnarray}
\label{lagrangian}
 L & = & -\frac{1}{4} G^{a}_{\mu\nu} G^{a\mu\nu} +
  \sum_{q=1}^{n_f} 
 \overline{\psi_q}\left( i D\!\!\!\!\slash\, - m_q \right)  \psi_q 
 +L_{gf}+L_{gc}
 \nonumber \\
 G_{\mu\nu}^{a}  & = & \partial_\mu A_\nu^{a} -\partial_\nu A_\mu^a
      + g f^{abc} A_\mu^b A_\nu^c  \nonumber \\
 \left[ D_{\mu}\right]_{ij} & = &
         \delta_{ij} \partial_\mu -i g A_\mu^a [T^a]_{ij}
\end{eqnarray}
The gauge-fixing and gauge-compensating parts of the Lagrangian
in the covariant gauge are
\begin{eqnarray}
L_{gf} & = & -\frac{1}{2\xi}(\partial^\mu A_\mu^a)^2 \nonumber \\ 
L_{gc} & = & \partial^{\mu}\overline{\omega}^a(\partial_\mu\omega^a
                 -g f^{abc} \omega^b A_\mu^c)
\end{eqnarray}
The fermion fields (the quark fields in QCD) $\psi_q$ have a 
mass $m_q$  and transform as the fundamental representation of
 a compact semi-simple Lie group,  $q=1,..., n_f$ is the flavor index.  
The Yang-Mills fields
(gluons in QCD) $A^a_\mu$ 
transform as the adjoint representation of this group.
$\omega^a$ are the ghost fields, and $\xi$ is the gauge parameter
of the covariant gauge.

$T^{a}$ are the generators of the fundamental representation
and  $f^{abc}$ are the structure constants of the Lie algebra,
\begin{equation}
T^aT^b-T^bT^a = i f^{a b c} T^c
\end{equation}
In the case of QCD we have the gauge group SU(3) but we will perform 
the calculation for an arbitrary compact semi-simple Lie group $G$.


The beta function and the quark mass anomalous dimension
are defined as 
\begin{eqnarray}
\label{beta}
\frac{d a }{d \ln \mu^2}  & = &
\beta(a) \nonumber \\
& = & -\beta_0 a^2 - \beta_1 a^3
-\beta_2 a^4
-\beta_3 a^5 + O(a^6) \hspace{1cm}
\end{eqnarray}
\begin{eqnarray}
\label{gamma}
\frac{d\ln m_q }{d \ln \mu^2}  & = &
\gamma_m(a) \nonumber \\
& = & -\gamma_0 a - \gamma_1 a^2
-\gamma_2 a^3
-\gamma_3 a^4 + O(a^5) \hspace{1cm}
\end{eqnarray}
where $m_q=m_q(\mu^2)$ is the renormalized (running) quark mass and
$\mu$ is  the renormalization point in the $\overline{\rm MS}$ scheme.
$a=\alpha_s/\pi=g^2/4\pi^2$ where $g=g(\mu^2)$ is the renormalized strong
coupling constant of the standard QCD Lagrangian of Eq. (\ref{lagrangian}).

This normalization for the basic perturbative expansion parameter $a$ 
 differs from the normalization $g^2/16\pi^2$ which    
naturally appears in multiloop calculations and was therefore used 
in \cite{beta4loop}.
Here we adopt the different normalization $a=\alpha_s/\pi$
because it is the commonly accepted normalization in applications. 

The 4-loop QCD $\beta$-function in the $\overline{\rm MS}$ scheme reads

\begin{eqnarray}
\renewcommand{\arraystretch}{ 1.3}
 \beta_0 & = & \frac{1}{4}\left[ 11 - \frac{2}{3} n_f \right]   \nonumber\\
\beta_1 & = & \frac{1}{16} \left[102 - \frac{38}{3} n_f \right] \nonumber \\
 \beta_2 & = & \frac{1}{64} \left[ \frac{2857}{2} - \frac{5033}{18} n_f 
   + \frac{325}{54}n_f^2 \right] \nonumber \\
 \beta_3 & = & \frac{1}{256} \left[  \left( \frac{149753}{6} 
            + 3564 \zeta_3 \right)
        - \left( \frac{1078361}{162} + \frac{6508}{27} \zeta_3 \right) n_f
 \right.  \nonumber \\ & &
  \left. + \left( \frac{50065}{162} + \frac{6472}{81} \zeta_3 \right) n_f^2
       +  \frac{1093}{729}  n_f^3 \right]
\end{eqnarray}
Or in a numerical form
\begin{eqnarray}
 \beta_0 & \approx & 2.75 - 0.166667  n_f \nonumber \\
 \beta_1 & \approx & 6.375 - 0.791667 n_f \nonumber \\
 \beta_2 & \approx & 22.3203 - 4.36892 n_f + 0.0940394 n_f^2  \nonumber \\
 \beta_3 & \approx & 114.23  -27.1339 n_f + 1.58238 n_f^2 + 0.0058567 n_f^3
\end{eqnarray}

To calculate the quark mass anomalous dimension $\gamma_m$ we need to calculate 
the renormalization constant $Z_m$ of the quark mass
\[ m_{B} = Z_m m\]
where $m_{B}$ is the bare (unrenormalized) quark mass.
Since the anomalous dimension $\gamma_m$ does not depend on masses
within the MS scheme it is formally the same for all quarks.
We will therefore omit the flavor index $q$ in $m_q$ when it is irrelevant.
The expression of the $\gamma_m$-function via $Z_m$ is given 
by the following chain of equations

\[ \frac{ d \ln m_{B}}{d \ln \mu^2} = 0 
   =  \frac{d \ln Z_m}{d \ln \mu^2 }
        + \frac{d \ln m}{d \ln \mu^2}  
\]
\begin{equation}
 \Rightarrow
 \gamma_m
 = 
 - \frac{d \ln Z_m}{d \ln \mu^2}=-\frac{\partial \ln Z_m}{\partial a}
       \frac{d a}{d \ln \mu^2 }
  =
-\frac{\partial \ln Z_m}{\partial a}\left[
  -\varepsilon a + \beta (a)\right]
 = a \frac{\partial Z_m^{(1)}}{\partial a} 
\end{equation}
where one uses the fact that the bare mass
 $m_{B}$ is invariant under the renormalization group
transformations.
Here $d a/(d \ln \mu^2) = -\varepsilon a + \beta (a)$ 
is the $\beta$-function in $D$-dimensions, 
$Z_{m}^{(1)}$ is the coefficient of the first 
$\varepsilon$-pole in $Z_{m}$ defined below.

Renormalization constants within the MS-scheme do not
depend on dimensional parameters (masses, momenta) \cite{collins} and
have the following structure:
\begin{eqnarray}
\label{zet}
Z_{m}(a,\frac{1}{\varepsilon})
=1+\sum_{n=1}^{\infty}\frac{Z_{m}^{(n)}(a)}{\varepsilon^n},
\end{eqnarray}
Since $Z_{m}$ does not depend explicitly on $\mu$ and $m$, 
the $\gamma_m$-function
is the same in all MS-like schemes, i.e. within the class of 
renormalization schemes which differ by the shift of the parameter $\mu$.
That is why the $\gamma_m$-function is the same in the MS-scheme \cite{h}
and in the $\overline{\rm MS}$-scheme \cite{bbdm}.

We obtained the mass renormalization constant $Z_m$ by 
calculating the following two renormalization constants of the Lagrangian:
$Z_{\overline{\psi}\psi}$ of the bilocal operator $\overline{\psi}\psi$
and $Z_{\psi}$ of the quark field (i.e. of the inverted quark propagator).
The connection between renormalized and bare quantities is defined as follows
\begin{eqnarray}
 [\overline{\psi}\psi]_R & = & \frac{Z_{\overline{\psi}\psi}}{Z_{\psi}}
  \overline{\psi}_{B}\psi_B \nonumber \\
  \psi_B & = & Z_{\psi}^{1/2} \psi
\end{eqnarray}
Then $Z_m=Z_{\overline{\psi}\psi} /Z_{\psi}$.
In order to calculate the 4-loop approximation of $Z_m$ we needed to
calculate the 4-loop ultraviolet counterterms of the quark
propagator and of the Green function
\begin{equation} G_{\overline{\psi}
[\overline{\psi}\psi ]\psi} =
 \int dx dy e^{i qx + ipy} 
  \langle 0 | T\{ \overline{\psi}(x) 
[\overline{\psi}(y)\psi(y)]  \psi(0) \} |0 \rangle
\end{equation}

These 4-loop ultraviolet counterterms were calculated using a technique which
is based on the method of infrared rearrangement \cite{vladimirov}
and is described in Ref.~\cite{beta4loop}.
This general technique reduces the calculation
of the counterterms to the direct calculation of 4-loop massive 
vacuum (bubble) integrals and provides a procedure that is well suited for the
automatic evaluation of large numbers of Feynman diagrams.
The calculations were done with the symbolic manipulation program 
FORM~\cite{form} and a specially designed database program that stores 
the results of the individual diagrams and adds them in the end.
For the present calculation we needed to evaluate of the order of
10000 4-loop diagrams. 
These diagrams were generated with the program QGRAF \cite{qgraph}.

We obtained in this way the following result for the 
4-loop $\gamma_m$-function in the $\overline{\rm MS}$-scheme
\renewcommand{\arraystretch}{1.3}
\begin{eqnarray} 
\label{eq:beta3}
\gamma_{0} & = & \frac{1}{4}\left[ 3 C_F \right]
 \nonumber \\
 \gamma_{1} & = & \frac{1}{16} \left[
 \frac{3}{2}C_F^2+\frac{97}{6} C_F C_A -\frac{10}{3} C_F T_F n_f 
  \right] \nonumber \\
 \gamma_{2} & = & \frac{1}{64} \left[
  \frac{129}{2} C_F^3 - \frac{129}{4}C_F^2 C_A
 + \frac{11413}{108}C_F C_A^2 \right. \nonumber \\ & & \left.
 +C_F^2 T_F n_f (-46+48\zeta_3)
+C_F C_A T_F n_f \left( -\frac{556}{27}-48\zeta_3 \right)  
- \frac{140}{27} C_F T_F^2 n_f^2  \right]  
   \nonumber \\
\gamma_{3} & = & \frac{1}{256} \left[
  C_F^4 \left(-\frac{1261}{8} - 336\zeta_3 \right)
   + C_F^3 C_A \left( \frac{ 15349}{12} + 316 \zeta_3 \right)
 \right. \nonumber \\ & & 
   + C_F^2 C_A^2 \left(-\frac{ 34045}{36} - 152 \zeta_3 + 440\zeta_5 \right)
   + C_F C_A^3 \left( \frac{70055}{72} + \frac{1418}{9} \zeta_3
                      - 440 \zeta_5 \right)
 \nonumber \\ & &
  + C_F^3 T_F n_f \left( -\frac{280}{3} + 552 \zeta_3 - 480 \zeta_5 \right)
  + C_F^2 C_A T_F n_f \left(- \frac{8819}{27} + 368 \zeta_3 
                            - 264 \zeta_4 + 80 \zeta_5 \right)
 \nonumber \\ & &
  + C_F C_A^2 T_F n_f \left(- \frac{65459}{162} 
                  - \frac{2684}{3} \zeta_3 + 264 \zeta_4
                    + 400 \zeta_5 \right)
 \nonumber \\ & &
  + C_F^2 T_F^2 n_f^2 \left( \frac{304}{27} - 160 \zeta_3 
                            + 96 \zeta_4 \right)
 \nonumber \\ & &
  + C_F C_A T_F^2 n_f^2 \left( \frac{1342}{81} 
                             + 160 \zeta_3 - 96 \zeta_4 \right) 
  + C_F T_F^3 n_f^3 \left(- \frac{664}{81} + \frac{128}{9} \zeta_3 \right)
 \nonumber \\ & & \left.
  + \frac{d_F^{a b c d}d_A^{a b c d}}{N_F}   
            \left(- 32 + 240 \zeta_3  \right)
  +  n_f \frac{d_F^{a b c d}d_F^{a b c d}}{N_F}   
            \left( 64 - 480 \zeta_3  \right) \right]
 \label{maingamma} \end{eqnarray}
Here $\zeta$ is the Riemann zeta function ($\zeta_3 = 1.2020569\cdots$,
$\zeta_4 = 1.0823232\cdots$ and $\zeta_5 = 1.0369277\cdots $).
 $[T^a T^a]_{ij} = C_F \delta_{ij}$ and
 $f^{a c d} f^{b c d} = C_A \delta^{ab}$ are the quadratic Casimir 
operators of the
fundamental and the adjoint representation of the Lie algebra and 
tr$(T^a T^b) = T_F \delta^{a b}$ is the trace normalization of the
fundamental representation. $N_F$ is the dimension of the fermion 
representation (i.e. the number of quark colours) and $n_f$ is the number of
 quark flavors. We expressed the higher order 
group invariants in terms of contractions between the following fully 
symmetrical tensors:
\begin{eqnarray}
 d_F^{a b c d} & = & \frac{1}{6 } {\rm Tr }  \left[
   T^a T^b T^c T^d
 + T^a T^b T^d T^c 
 + T^a T^c T^b T^d  \right. \nonumber \\
 & & \left. \hspace{4mm}
 + T^a T^c T^d T^b 
 + T^a T^d T^b T^c 
 + T^a T^d T^c T^b  \hspace{1mm}
 \right]\\
\label{eq:CCCC}
 d_A^{a b c d} & = & \frac{1}{6} {\rm Tr }  \left[
   C^a C^b C^c C^d
 + C^a C^b C^d C^c 
 + C^a C^c C^b C^d  \right. \nonumber \\
 & & \left. \hspace{4mm}
 + C^a C^c C^d C^b 
 + C^a C^d C^b C^c 
 + C^a C^d C^c C^b  \hspace{1mm}
 \right]
\end{eqnarray}
where the matrices $[C^a]_{bc} \equiv - i f^{abc}$ are the generators in the
adjoint representation. 

The result of Eq. (\ref{maingamma}) 
is valid for an arbitrary semi-simple compact Lie group. 
The result for QED (i.e. the group U(1))
is included in Eq. (\ref{maingamma}) by substituting
$C_A = 0$, $d_A^{a b c d} = 0$, $C_F = 1$, $T_F = 1$, $(d_F^{a b c d})^2=1$,
$N_F = 1$.
The $n_f^3$ and $n_f^2$ terms in this 4-loop result for QED agree with the
literature \cite{gammaQEDlargenf}
where the leading and next-to-leading large-$n_f$ terms for
the QED gamma-function were calculated in all orders of the coupling constant.
Furthermore it is interesting to note that for the choice
$C_F = C_A = T_F, n_f = 1/2$ and $d_F = d_A$, which
corresponds to the case of N=1 supersymmetry,
all $\zeta$ terms and terms with the tensors
$d$ cancel. This is analogous to the case of the 4-loop
$\beta$-function \cite{beta4loop}.

The result of Eq.(\ref{maingamma}) was obtained in an arbitrary
covariant gauge for the gluon field.
This means that we have kept 
the gauge parameter $\xi$ that appears in the gluon propagator
$ i [-g^{\mu\nu}+(1-\xi)
 q^{\mu}q^{\nu}/(q^2+i\epsilon)]/(q^2+i\epsilon)$  as a
free parameter in the calculations. The explicit cancellation of the gauge
dependence in $\gamma_m$ gives an important check of the results.
The results for individual
diagrams that contribute to $\gamma_m$ also contain 
several constants specific for massive vacuum integrals.
The cancellation of these constants at various stages in the calculation
provides additional checks of the result. 
We note that at the 4-loop level $\zeta_4$ and $\zeta_5$ 
(Riemann zeta function of arguments 4 and 5) appear as new constants.

For the standard normalization of the SU($N$) generators we find
the following expressions for the color factors
(more details about the color factors for various other groups can be found 
in \cite{beta4loop})

\[ T_F = \frac{1}{2}, \hspace{.5cm} N_F = N,  \hspace{.5cm} C_A = N, 
\hspace{.5cm} C_F = \frac{N^2-1}{2 N} , \]
\[  \frac{d_F^{a b c d}d_A^{a b c d}}{N_F}  =
                          \frac{ (N^2-1)(N^2+6)}{48}, \hspace{.5cm}
  \frac{d_F^{a b c d}d_F^{a b c d}}{N_F}  =
                  \frac{(N^2-1)(N^4-6N^2+18)}{96 N^3} \]
Substitution of these color factors for $N=3$ into 
Eq. (\ref{maingamma}) yields
the following result for QCD
\begin{eqnarray}
\renewcommand{\arraystretch}{ 1.3}
\label{eq:gamma}
 \gamma_0 & = & 1   \nonumber\\
\gamma_1 & = & \frac{1}{16}\left[ \frac{202}{3}
                 - \frac{20}{9} n_f \right]       
  \nonumber \\
 \gamma_2 & = & \frac{1}{64} \left[1249+\left( - \frac{2216}{27} 
          - \frac{160}{3}\zeta_3 \right)n_f 
               - \frac{140}{81} n_f^2 \right]
\nonumber \\
 \gamma_3 & = & \frac{1}{256} \left[ 
       \frac{4603055}{162} + \frac{135680}{27}\zeta_3 - 8800\zeta_5
 +\left(- \frac{91723}{27} - \frac{34192}{9}\zeta_3 
    + 880\zeta_4 + \frac{18400}{9}\zeta_5 \right) n_f
 \right.  \nonumber \\ & & \left.
 +\left( \frac{5242}{243} + \frac{800}{9}\zeta_3 
    - \frac{160}{3}\zeta_4 \right) n_f^2
 +\left(- \frac{332}{243} + \frac{64}{27}\zeta_3 \right) n_f^3 \right]
\end{eqnarray}
Or in a numerical form
\begin{eqnarray}
 \gamma_0 & = & 1 \nonumber \\
 \gamma_1 & \approx & 4.20833 - 0.138889 n_f  \nonumber \\
 \gamma_2 & \approx & 19.5156 - 2.28412 n_f - 0.0270062 n_f^2  \nonumber \\
 \gamma_3 & \approx & 98.9434 -19.1075 n_f+ 0.276163 n_f^2
                       +0.00579322 n_f^3 
\end{eqnarray}

Let us now consider the solution of the evolution (i.e. renormalization 
group) equation Eq.(\ref{gamma}) for the  quark mass
\begin{equation}
\label{solrg}
 m_q(\mu^2)= m_q(\mu_0^2)
  \exp \left( \int_{a(\mu_0^2)}^{a(\mu^2)}
        da' \frac{\gamma_m(a')}{\beta(a')}\right)
       ={\hat m}_q
      \exp \left( \int^{a(\mu^2)}
        da' \frac{\gamma_m(a')}{\beta(a')}\right)
\end{equation}
where we formally define the renormalization group invariant (i.e. independent
of $\mu^2$) quark mass ${\hat m}_q$  as
\begin{equation}
\label{minv}
 {\hat m}_q = m_q(\mu_0^2)
  \exp \left(- \int^{a(\mu_0^2)}
        da' \frac{\gamma_m(a')}{\beta(a')}\right)
\end{equation}
One can expand Eq.(\ref{solrg}) in $a$ to obtain the  following
perturbative solution to the evolution equation
\begin{equation}
\label{expand}
m_q(\mu^2)={\hat m}_q a^{\gamma_0/\beta_0}
 \left[1+A_1~ a +\left(A_1^2+A_2 \right) \frac{a^2}{2}
+\left( \frac{1}{2}A_1^3+\frac{3}{2} A_1 A_2 +A_3 \right) \frac{a^3}{ 3}
 +O(a^4) \right]
\end{equation}
where

\begin{eqnarray}
A_1 & = & -\frac{\beta_1 \gamma_0}{\beta_0^2}+\frac{\gamma_1}{\beta_0}
 \nonumber \\
A_2 & = &  \frac{\gamma_0}{\beta_0^2}
   \left(\frac{\beta_1^2}{\beta_0}-\beta_2\right)
  -\frac{\beta_1 \gamma_1}{\beta_0^2}+\frac{\gamma_2}{\beta_0}
 \nonumber \\
A_3 & = & \left[\frac{\beta_1 \beta_2}{\beta_0}
  -\frac{\beta_1}{\beta_0} 
     \left(\frac{\beta_1^2}{\beta_0}-\beta_2 \right)
          -\beta_3 \right] \frac{\gamma_0}{\beta_0^2}
   +\frac{\gamma_1}{\beta_0^2} \left(\frac{\beta_1^2}{\beta_0}-\beta_2 \right) 
   -\frac{\beta_1 \gamma_2}{\beta_0^2}
   +\frac{\gamma_3}{\beta_0}
 \nonumber
\end{eqnarray}

This gives us the following 4-loop
expansions for the running $\overline{\rm MS}$ quark masses $m_q(\mu^2)$
for quark flavors q=s,c,b,t
\begin{eqnarray}
\label{evolution}
m_s(\mu^2) & = & {\hat m}_s \left(\frac{\alpha_s}{\pi}\right)^{4/9}
~ \left[1 + 0.895062 \left(\frac{\alpha_s}{\pi}\right)
 + 1.37143 \left(\frac{\alpha_s}{\pi}\right)^2 
 + 1.95168 \left(\frac{\alpha_s}{\pi}\right)^3  \right]
\nonumber \\
m_c(\mu^2) & = & {\hat m}_c \left(\frac{\alpha_s}{\pi}\right)^{12/25}
 \left[1 + 1.01413 \left(\frac{\alpha_s}{\pi}\right)
 + 1.38921 \left(\frac{\alpha_s}{\pi}\right)^2  
 + 1.09054 \left(\frac{\alpha_s}{\pi}\right)^3   \right]
\nonumber \\
m_b(\mu^2) & = & {\hat m}_b \left(\frac{\alpha_s}{\pi}\right)^{12/23}
 \left[1 + 1.17549 \left(\frac{\alpha_s}{\pi}\right)
+ 1.50071 \left(\frac{\alpha_s}{\pi}\right)^2  
+ 0.172478 \left(\frac{\alpha_s}{\pi}\right)^3 \right]
\nonumber \\
m_t(\mu^2) & = & {\hat m}_t \left(\frac{\alpha_s}{\pi}\right)^{4/7}
~ \left[1 + 1.39796 \left(\frac{\alpha_s}{\pi}\right)
+ 1.79348 \left(\frac{\alpha_s}{\pi}\right)^2  
- 0.683433 \left(\frac{\alpha_s}{\pi}\right)^3  \right]
\end{eqnarray}
where $\alpha_s \equiv \alpha_s(\mu^2/\Lambda_{\overline{\rm MS}}^2)$.

In phenomenological applications one uses the $\overline{\rm MS}$
quark masses at a characteristic scale $\mu$ of a considered process 
to sum large logarithms.
From the expansions (\ref{evolution})
we conclude that the invariant mass ${\hat m}_q$ is
a good reference mass for the accurate evolution of the 
$\overline{\rm MS}$ quark masses to the necessary scale $\mu$
in phenomenological applications.
The perturbative coefficients of the solutions (\ref{evolution})
of evolution equations
for the $\overline{\rm MS}$ quark masses are small 
up to and including the 4-loop level.
On the contrary, it is known that the equations expressing
the $\overline{\rm MS}$ quark masses $m_q$ via the pole quark masses $M_q$
have worse convergence of the perturbative expansions \cite{gbgs};
for a review see Ref.~\cite{ckk}.
We should emphasize in this respect that the invariant mass is a fundamental
gauge-invariant object that naturally appears in the solution of the 
renormalization group equation.

\section*{The analysis of the QCD infrared fixed point}
Let us apply the obtained 4-loop quark mass anomalous dimension
to the analysis of the QCD infrared fixed point in the case
of the Higgs boson decay into hadrons.
The infrared fixed point was analyzed at the third-order of
perturbative QCD in Ref.~\cite{gkls} for the total hadronic Higgs decay
and in \cite{ckl,ms} for electron-positron annihilation into hadrons. 
Recently the total hadronic decay width of the Higgs boson was calculated
at the 4-loop level of perturbative QCD~\cite{chet} in the limit
of massless quarks. 
\begin{eqnarray}
\label{higgs4loop}
\Gamma_H & = &
 \frac{3G_F}{4\sqrt{2}\pi} M_H \sum_{q} m_q^2
   \Biggl\{ \Biggr. 1 + \frac{17}{3} a
  +  a^2 \Biggl[ \Biggr. 
 \frac{10801}{144} -\frac{19}{12} \pi^2-\frac{39}{2}\zeta_3 
\nonumber \\ & &
 + n_f \left(-\frac{65}{24}+\frac{1}{18}\pi^2  +\frac{2}{3}\zeta_3 \right)
       \Biggl.  \Biggr]
  +  a^3 \Biggl[ \Biggr.
\frac{6163613}{5184} -\frac{3535}{72}  \pi^2
-\frac{109735}{216} \zeta_3
+\frac{815}{12} \zeta_5 
 \nonumber \\ & &
  +  n_f \left(-\frac{46147}{486} +\frac{277}{72}\pi^2
+ \frac{262}{9}\zeta_3 -\frac{5}{6}\zeta_4 -\frac{25}{9}\zeta_5 \right)
  +  n_f^2 
\left( \frac{15511}{11664} -\frac{11}{162}  \pi^2 -\frac{1}{3}\zeta_3 \right)
              \Biggl.  \Biggr]    \Biggl. \Biggr\}
\nonumber \\ &
\approx & \frac{3G_F}{4\sqrt{2}\pi}M_H \sum_{q} m_q^2 \left[ \right.
 1 + 5.66667 a
+ \left(35.93996 - 1.35865 n_f\right) a^2 
\nonumber \\ & &
+
\left(164.1392 - 25.77119 n_f + 0.258974 n_f^2\right)a^3
\left. \right]
\end{eqnarray}
where $ a = \frac{\alpha_s(M_H)}{\pi}, m_q=m_q(M_H)$ and $M_H$ is
the Higgs mass. We should note that if the Higgs boson is lighter than
the top quark, then additional contributions appear
from the so-called singlet diagrams due to
non-decoupling of the heavy top quark in this channel
\cite{ck}, see also Ref.~\cite{higlrv}. 
We will neglect these contributions.

Together with the 4-loop quark mass anomalous dimension it allows
for the first time to perform the analysis of the 
physical quantity at the fourth-order of 
perturbative QCD, i.e. with {\em four} known perturbative QCD 
$\alpha_s$-terms.

For the analysis of the infrared fixed point 
it is convenient instead of $\Gamma_H$
to introduce the function
\[
R(a)=- \frac{1}{2}\frac{ d \ln(\Gamma_H/M_H)}{d \ln M_H^2}= 
 -\gamma_m(a)-\frac{1}{2}\beta(a)\frac{\partial \ln 
\Gamma(a)}{\partial a}
\]
\begin{equation}
\label{rs}
=r_0a(1+r_1 a +r_2 a^2 +r_3 a^3) +O(a^5)
\end{equation}
where $ \Gamma_H
  \equiv  \frac{3G_F}{4\sqrt{2}\pi} M_H \sum_{q} m_q^2 \Gamma(a)$
with $\Gamma(a) \equiv 
1+\Gamma_1 a+\Gamma_2 a^2 +\Gamma_3 a^3$. This gives
\begin{eqnarray}
\label{ri}
r_0 & = & \gamma_0 =1 
\nonumber \\
r_1 & = & \gamma_1+\frac{1}{2}\beta_0 \Gamma_1
\nonumber \\
r_2 & = & \gamma_2 +\frac{1}{2}(\beta_1 \Gamma_1
-\beta_0 \Gamma_1^2+2 \beta_0 \Gamma_2)
\nonumber \\
r_3 & = & \gamma_3+
\frac{1}{2}(\beta_0\Gamma_1^3- \beta_1\Gamma_1^2+ \beta_2\Gamma_1
   +2 \beta_1 \Gamma_2
        -3 \beta_0 \Gamma_1\Gamma_2+3 \beta_0\Gamma_3)
\end{eqnarray}
And the full result for $R$ becomes
\begin{eqnarray}
\label{rires}
r_1 & = & 12 -\frac{11}{18}n_f 
\nonumber \\ & \approx & 12 - 0.61111 n_f \\
\nonumber \\
r_2 & = & \frac{7189}{36} -\frac{209}{48}\pi^2-\frac{429}{8}\zeta_3
 +n_f \left(-\frac{2995}{144}+\frac{5}{12}\pi^2+\frac{17}{4}\zeta_3\right)
\nonumber \\ & &
 +n_f^2 \left(\frac{275}{648}-\frac{1}{108}\pi^2 -\frac{1}{9}\zeta_3\right)
\nonumber \\ &
\approx & 92.26 - 11.578 n_f + 0.19944 n_f^2 \\
\nonumber \\ 
r_3 & = & \frac{81937369}{20736}-\frac{11239}{64}\pi^2
 -\frac{3014503}{1728}\zeta_3+\frac{7865}{32}\zeta_5
\nonumber \\ & &
+n_f\left(-\frac{ 4313797}{6912}+\frac{15097}{576}\pi^2
 +\frac{180343}{ 864}\zeta_3-\frac{2945}{144}\zeta_5\right)
\nonumber \\ & &
+n_f^2\left(\frac{1734463}{62208}-\frac{1043}{864}\pi^2
-\frac{71}{9}\zeta_3+\frac{25}{36}\zeta_5\right)
+n_f^3\left(-\frac{ 985 }{2916}+\frac{11}{648}\pi^2+\frac{5}{54}\zeta_3
\right)
\nonumber \\ &
\approx & 376.12 - 135.72 n_f + 7.2045 n_f^2  - 0.05895 n_f^3
\end{eqnarray} 
One may notice that the $\zeta_4$-terms coming separately from the 4-loop 
mass anomalous dimension and the 4-loop decay width cancel in the final 
expression for $R(a)$. This cancellation of $\zeta_4$-terms 
provides an extra cross-check for both the calculations of $\gamma_m$ and
$\Gamma_H$.
The remaining $\pi^2$-terms come from the imaginary parts of logarithms in the 
process of the analytical continuation from the Euclidean to the Minkowski 
region.

For the analysis of the infrared fixed 
point we will apply the approach of effective
charges \cite{grunberg}. This approach to the 
scheme dependence of perturbative series
is known, see e.g. Ref.~\cite{ckl}, to give values for physical
quantities close to another distinguished renormalization scheme
-- optimized perturbation theory \cite{stev}.
Within the approach of effective charges one defines the whole
calculated perturbative expansion as a new effective charge, i.e.
\begin{equation}
\label{defef}
R(a)=a+r_1a^2+r_2a^3+r_3a^4\equiv a_{\rm eff}
\end{equation}
The evolution equation for the effective charge (or equivalently
for the physical quantity itself in the given order of
perturbation theory) is governed
by an effective $\beta$-function 
\begin{eqnarray}
\label{effevol}
\frac{d R }{d \ln M_H^2}  & = &
\beta_{\rm eff}(R) \nonumber \\
& = & -\beta_0 R^2-\beta_1 R^3
 -\beta_2^{\rm eff} R^4 -\beta_3^{\rm eff} R^5 + O(R^6)
\end{eqnarray}
Thus in the effective charge approach one deals with only one
asymptotic perturbative series for $\beta_{\rm eff}(R)$ instead of two
series for $R(a)$ and for $\beta(a)$ in the $\overline{\rm MS}$-scheme.  

The third and the fourth coefficients of the effective $\beta$-function
(which are scheme-invariant since they govern the evolution of the
physical quantity $R(a)$)
are expressed via the $\overline{\rm MS}$ coefficients as follows

\begin{eqnarray}
\label{rhos}
\beta_2^{\rm eff} 
 & = & \beta_2 -\beta_1 r_1+\beta_0 (r_2 -r_1^2)
\nonumber \\
\beta_3^{\rm eff} & = & \beta_3 -2\beta_2 r_1+\beta_1 r_1^2 
+2 \beta_0 (r_3 -3 r_1r_2 +2 r_1^3) 
\end{eqnarray}
The numerical results are
\begin{eqnarray}
\label{betaeff}
 \beta_2^{\rm eff} 
 & \approx &  -196.464 + 26.1453 n_f - 1.38317 n_f^2  + 0.0290035 n_f^3
\\
\nonumber \\
\beta_3^{\rm eff} 
 & \approx & 3305.698 - 700.571 n_f + 65.1914 n_f^2  - 2.89465 n_f^3 
 + 0.0499219 n_f^4
\label{beta3eff}
\end{eqnarray}

In the third order of QCD this $\beta_{\rm eff}$-function has a positive zero
$R^{(0)}\approx 0.15$ \cite{gkls} which practically 
does not depend on the number
of quark flavors for $n_f=3,4,5,6$. 
In Ref.~\cite{gkls} it was considered to be a spurious fixed point, but 
it could have been interpreted as a real QCD infrared fixed point 
which would indicate the possibility
of applying perturbation theory till zero energy. This is why
it is important to check the stability of this zero $R^{(0)}$ under
the inclusion of the fourth order of perturbative QCD.
One can see from Eq.~(\ref{beta3eff}) that at the fourth order level 
the positive zero of the effective beta function disappears. 
This supports the idea that the infrared fixed point for the Higgs boson
decay into hadrons at the third order of perturbative QCD is indeed
a spurious fixed point.


\section*{Acknowledgements}
We are grateful to J.A. Gracey, P.J. Nogueira and A.N. Schellekens for 
helpful and stimulating discussions. S.L. is grateful to the 
Theory Group of NIKHEF and the Theory Division of CERN for 
their kind hospitality at different stages of the work; 
his work is supported in part by the Russian 
Foundation for Basic Research grant 96-01-01860. The work of T.R. is 
supported by the US Department of Energy.

Just before completing this paper a similar paper appeared~\cite{Chetalfa4}
with the mass anomalous dimension to four loops. It agrees completely with
our formulae~\ref{eq:gamma} and~\ref{evolution}.

\end{document}